\documentclass[letterpaper,11pt]{article}
\pdfoutput=1

\usepackage{jheppub}
\usepackage{booktabs} 
\usepackage{graphicx} 
\usepackage{siunitx}
\usepackage{cleveref}
\usepackage{epsfig}
\usepackage[normalem]{ulem}
\usepackage{bm}
\usepackage{bbm}
\usepackage{slashed}
\usepackage{xspace}
\usepackage{subfigure}
\usepackage{multirow}

\usepackage{amsmath,latexsym}
\usepackage{pstricks}
\usepackage{color} 
 
\usepackage{marginnote}

\newcommand{\nn}{\nonumber}

\newcommand{\beq}{\begin{equation}}
\newcommand{\eeq}{\end{equation}}
\newcommand{\bea}{\begin{eqnarray}}
\newcommand{\eea}{\end{eqnarray}}

\begin{document}


\title{Horizon radiation reaction forces}

\author[a]{Walter D. Goldberger,}
\author[b]{Ira Z. Rothstein}
\affiliation[a]{Department of Physics, Yale University, New Haven, CT 06511}
\affiliation[b]{Department of Physics, Carnegie Mellon University, Pittsburgh, PA 15213}

\vspace{0.3cm}


\abstract{
Using Effective Field Theory (EFT) methods, we compute the effects of horizon dissipation on the gravitational interactions of relativistic binary black hole systems.    We assume that the dynamics is perturbative, i.e it admits an expansion in powers of Newton's constant (post-Minkowskian, or PM, approximation).    As applications, we compute corrections to the scattering angle in a black hole collision due to dissipative effects to leading PM order,  as well as the post-Newtonian (PN) corrections to the equations of motion of binary black holes in non-relativistic orbits, which represents the leading order finite size effect in the equations of motion.   The methods developed here are also applicable to the case of more general compact objects, eg. neutron stars, where the magnitude of the dissipative effects depends on non-gravitational physics (e.g, the equation of state for nuclear matter).
 }
\maketitle



\section{Introduction}

Gravitational wave measurements~\cite{exp} have the potential to extract information about the internal dynamics of compact astrophysical objects.
For the early stages of a binary inspiral these effects are relatively small.   In particular, conservative finite size (i.e. tidal) effects scale as the fifth power of the object's radius,  and enter formally at fifth order in the Post-Newtonian (5PN) expansion for non-relativistic binary dynamics.  Although these tidal corrections are known to vanish for Schwarzschild black holes~\cite{Binnington:2009bb,Damour:2009va,Kol:2011vg}, for neutron stars the effects can be somewhat enhanced~\cite{Flanagan:2007ix,Hinderer:2007mb} due to the fact that the physical radius is larger than the gravitational radius $r_s= 2 G_N M$.

In this paper we focus on the relatively less well-understood effects of \emph{dissipation} in binary dynamics.  For the case of black hole binaries, this source of dissipation is the horizon itself, which absorbs energy-momentum as well as spin, and has a non-trivial effect on gravitational wave observables.   For non-spinning black holes, these effects were first studied in~\cite{Poisson:1994yf,Tagoshi:1997jy} for the case of a point mass in circular orbit around a much heavier black hole, and extended in~\cite{Poisson:2004cw} to study the effects of horizon dissipation for the motion of a small black hole in a fixed background spacetime whose curvature exceeds the Schwarzschild radius $r_s$.   More generally, one should expect absorptive effects due to the presence of low-lying internal modes (e.g. hydrodynamic fluctuations) that get excited during the evolution of the binary.

In ref.~\cite{GnR2}, we used the effective field theory formalism developed in \cite{GnR1,Goldberger:2006bd}, to include such dissipative effects in binary dynamics in a model independent fashion, by ``integrating back in'' the gapless degrees of freedom that are responsible for the dissipation (for the case of spinning object see \cite{Porto:2007qi}). Technically this is accomplished by fibering the worldline of the compact object with the Hilbert space\footnote{Since the formalism is quantum mechanical, it can also be used to capture the effects of Hawking radiation in the dynamics of interacting black holes.   See~\cite{Houses}.} of the underlying degrees of freedom. In the long distance, derivative, expansion we write down all possible couplings of the gravitational field to a collection of operators in a way which is constrained by diffeomorphism invariance.   The effect of these degrees of
freedom on the equations of motion of the binary are then given in terms of the correlation functions of these operators, which can be extracted from low energy processes (graviton absorption and emission) in the single body sector.

The original treatment  in ref.~\cite{GnR2} was based on the standard Dyson time-ordered formalism appropriate for the computation of $S$-matrix elements, and is therefore limited to capturing the effects of dissipation on time averaged observables, e.g. the power loss into internal degrees of freedom averaged over many orbital cycles.   In this paper, we extend the framework to include real time dynamics, in particular how absorption enters into the time evolution of the orbital motion, focusing here on the case of non-spinning black holes (the case of spin will be presented in forthcoming work).   To do this, we adopt the approach of ref.~\cite{galley}, which employed the Schwinger-Keldysh closed time path, or ``In-In'', formalism~\cite{Schwinger:1960qe,Keldysh:1964ud} in order to describe leading order PN radiation reaction effects~\cite{BT} in the context of EFT.   

In sec.~\ref{sec:inin}, we summarize our EFT setup.  To illustrate the formalism, we compute the  dissipative forces on a small Schwarzschild black hole that propagates in a fixed background gravitational field.   Given that the conservative part of the black hole quadrupolar response vanishes at zero frequency, our result captures the leading order deviation from geodesic motion due to  horizon effects.  The result given in Eq.~(\ref{eq:dpdsbh}), has not appeared previously in the literature.  However, it is consistent with results found in~\cite{Poisson:2004cw} upon time averaging over the black hole's worldline.   

In sec.~\ref{sec:2body}, we generalize to horizon absorption in gravitational dynamics of comparable mass black holes, in both PM and PN limits.   First, we consider large impact parameter \emph{inelastic} scattering\footnote{The complementary case of inelastic quantum mechanical scattering off a black hole is discussed in~\cite{GnR4}.} of two relativistic black holes induced by horizon absorption,neglecting the emission of gravitational radiation.   This is partly motivated by recent developements connecting classical PM scattering in general relativity (see e.g.~\cite{Westpfahl:1979gu,Portilla:1980uz,Bel:1981be,Damour:2016gwp,Damour:2017zjx,Vines:2017hyw,Bini:2018ywr,Vines:2018gqi,Antonelli:2019ytb,Siemonsen:2019dsu,Kalin:2019rwq,Kalin:2019inp,Damour:2019lcq,Bini:2020flp}), the Effective One Body (EOB) \cite{EOB} approach used to model the intermediate stages of the binary inspiral, and modern scattering amplitude approaches to classical gravitational scattering, see e.g.~\cite{Neill:2013wsa,Cachazo:2017jef,Cheung:2018wkq,Kosower:2018adc,Guevara:2018wpp,Bern:2019nnu,Bautista:2019sca,Cristofoli:2019neg,Maybee:2019jus,Guevara:2019fsj,Bern:2019crd,Bjerrum-Bohr:2019kec,Cristofoli:2020uzm,Cheung:2020sdj}.   Specifically, in sec.~\ref{sec:PM}, we obtain, for fixed initial data, the inelastic corrections to the PM momentum deflection during the collisions, and use this result to determine the corrections to the CM frame scattering angle (an 8PM effect), as well as the differential distribution (cross section) of final black hole masses in inelastic collisions.    The inelastic PM corrections we obtain are not easily accessible to scattering amplitude methods (based on an underlying unitary $S$-matrix description of point particles) and it would be interesting to incorporate our results in sec.~\ref{sec:PM} into the EOB formalism, as has been recently discussed in refs.~\cite{Damour:2016gwp,Damour:2017zjx,Bini:2018ywr,Antonelli:2019ytb,Damour:2019lcq,Bini:2020flp} in the context of elastic scattering. 

 In sec.~\ref{sec:PN} we compute, as a second application, the PN corrections to the equations of motion of a binary black hole system.   The resulting contribution to the gravitational force on each object, which is time-reversal odd, acts as a damping term at 6.5PN order, and is analogous to the Burke-Thorne~\cite{BT} 2.5PN potential induced by gravitational radiation reaction.   It represents the leading order finite size effect on the equations of motion for PN black hole binaries.   

Even though in this paper our focus is the inelastic dynamics of black holes, our methods can be straightforwardly generalized to the case of dissipation for binary inspirals of other compact objects.   We briefly comment on the magnitude of such effects and outline other extensions of the work presented here in sec.~\ref{sec:conc}.

\section{In-In formalism for black hole dissipative forces}
\label{sec:inin}

In this section, we consider how internal dissipative processes affect the orbital motion of a compact object moving through a fixed background spacetime.     Although the methods we discuss in this section can be used to treat a generic compact object that carries internal degrees of freedom, we will focus on the case of a small (non-spinning) black hole moving through a fixed background spacetime. We will generalize to the case of dynamical spacetimes (black holes binaries) in the next section.  

We work in the limit $r_s\ll {\cal R}$ in which the horizon's radius $r_s=2 G_N M$ is much smaller than the typical length scale ${\cal R}$ over which the background metric varies (the curvature radius).   In this case, the black hole may be described as an effective worldline, with finite size effects encapsulated by local, curvature dependent terms in a generalized point-particle Lagrangian.

In order to account for dissipative effects while retaining a point particle description, we use the formalism introduced in~\cite{GnR2}.   In this approach, dissipation of long wavelength gravitational energy by the compact object is attributed to the existence of gapless modes localized on the worldline which absorb energy as well as linear and angular momentum from the external environment.  Regardless of their microscopic origin, as we showed~\cite{GnR2}, these modes can be integrated back in  in a \emph{model independent fashion} in order to systematically account for dissipative effects of gravitationally interacting  compact objects.

In the absence of dissipative or finite size effects, the point particle action is 
\beq
\label{Spp}
S_{pp} = -\int dx^\mu  p_\mu + {1\over 2} \int d\lambda e \left(g^{\mu\nu} p_\mu p_\nu-m^2\right) 
\eeq 
We find it convenient to express the theory in Hamiltonian form, so that in addition to the trajectory $x^\mu(\lambda)$, we also introduce its conjugate momentum variable $p_\mu = -\delta S_{pp}/\delta {\dot x}^\mu$.    The ``einbein''  $e(\lambda)$ enforces worldline reparameterization invariance $\lambda\mapsto \lambda'(\lambda)$, $e'(\lambda') d\lambda' = e(\lambda) d\lambda$, so that the variable $s$, defined by 
\beq
ds = e(\lambda) d\lambda,
\eeq
is invariant under reparameterizations.    Varying the action with respect to the the kinematic variables $\chi(\lambda)=(x^\mu,p_\mu)$,  leads to the equations $p^\mu = dx^\mu/ds$,  $dp_\mu/ds = {1\over 2}\partial_\mu g_{\rho\sigma} p^\rho p^\sigma$, which together with the on-shell constraint $p^2 = m^2$ resulting from the variation of the einbein are equivalent to the usual geodesic equation for the trajectory $x^\mu(\lambda)$.

To include dissipation, we now introduce a set of internal worldline degrees of freedom, which we generically label as $X(\lambda)$ with reparametrization invariant action 
\beq
S_X = \int d\lambda e L_X(X,e^{-1} {\dot X})  - \int d\lambda e Q^E_{\mu\nu}(X,e) E^{\mu\nu}(x,p)  - \int d\lambda e Q^B_{\mu\nu}(X,e) B^{\mu\nu}(x,p) + \cdots.
\eeq
Here $L_X$ is some microscopic Lagrangian for the internal modes, whose form we will not need to know in order to obtain our results.   We have also explicitly written down the interaction of these modes with an external gravitational field, to leading order in gradients of the spacetime metric,   via the introduction  of dynamical moments $Q^E_{\mu\nu}$ and $Q^B_{\mu\nu}$,  which are `composite operators' built out of the variables $X(\lambda),e(\lambda)$ in some unspecified way.  They couple to the electric and magnetic components of the Weyl tensor $W_{\mu\nu\rho\sigma}$,
\bea
E_{\mu\nu} &=& W_{\mu\rho\nu\sigma} {p^\rho p^\sigma\over p^2} ,\\
B_{\mu\nu} &=& {\tilde W}_{\mu\rho\nu\sigma} {p^\rho p^\sigma\over p^2}= {1\over 2} \epsilon_{\mu\rho\lambda\gamma} W^{\rho \lambda}{}_{\nu\sigma} {p^\gamma p^\sigma\over p^2}. 
\eea
Note that we have not included any explicit couplings of $X(\lambda)$ to the Ricci curvature, as these can be removed by field redefinitions of the spacetime metric and therefore have no physical content.   For the same reason, only the traceless, transverse to $p^\mu$ components of the tensors $Q^{E,B}_{\mu\nu}$ couple to the external field.   The full dynamics of the compact object coupled to gravity is then encoded in the sum $S=S_{pp}(\chi,e)+S_X(X,e)$.

The objective now is to determine how the internal modes $X$ affect the time evolution of $x^\mu, p_\mu$.    We will treat the modes $X$ quantum mechanically, and since we are interested in real time dynamics, the correct formulation is the In-In (or Schwinger-Keldysh) closed time path integral~\cite{Schwinger:1960qe,Keldysh:1964ud}.   In this approach, we integrate out $X$ from a path integral with doubled variables, 
\beq
\label{eq:inin}
\exp\left[{i\Gamma[\chi,e;{\tilde\chi},{\tilde e}}]\right] = \int D X D {\tilde X} \exp\left[iS[\chi,X,e]-iS[{\tilde \chi},{\tilde X},{\tilde e}]\right]
\eeq
to obtain a functional $\Gamma[\chi,e;{\tilde \chi},{\tilde e}]$ (the In-In effective action) whose variation yields the classical motion of the kinematic variables $\chi=(x,p)$,
\beq
\left.{\delta\over \delta \chi(\lambda)}\Gamma[\chi,e;{\tilde\chi},{\tilde e}]\right|_{{\tilde \chi},{\tilde e}=\chi,e} =0.
\eeq
Note that by construction, the In-In action vanishes when we set ${\tilde\chi}=\chi,{\tilde e}=e$.

The variation with respect to the einbein $e(\lambda)$  generates a mass shell condition relating the invariant mass $p^2=g_{\mu\nu} p^\mu p^\nu $ to the internal degrees of freedom,
\beq
p^2 = m^2 + 2 \langle H_X + H_{int}\rangle.
\eeq
This equation reflects the transfer of energy between the kinematic (orbital) modes and the internal degrees of freedom $X$ as the object propagates through a tidal environment.    Here, $\langle\cdots\rangle$ denotes a quantum expectation value in the initial state of the internal modes $X(\lambda)$, and correspond to the In-In path integral expression
\beq
\langle {\cal O}[X]\rangle = \int D X D {\tilde X}  e^{iS[\chi,e,X]-iS[{\chi},e {\tilde X}]} {\cal O}[X]
\eeq
for any composite operator ${\cal O}$.  In general, this expectation  value depends on the worldline variables $(x,p,e)$ as well as any other external fields that couple to the particle.     The internal Hamiltonian in the absence of interaction is 
\beq
H_X = -{\delta\over\delta e} \int d\lambda e L_X(X,e^{-1} {\dot X}) = {\dot X} {\partial L_X\over \partial {\dot X}} - L_X,
\eeq
while the tidal coupling gives
\beq
H_{int} = {\delta\over\delta e}\int d\lambda e \left(Q^E_{\mu\nu} E^{\mu\nu}+Q^B_{\mu\nu} B^{\mu\nu}\right).
\eeq

Below, we will use this formalism to calculate the effects of dissipation on the dynamics of gravitationally interacting compact objects, focusing on the case of Schwarzschild black holes.  First, consider as a simplifying case a small black hole propagating in a fixed background $g_{\mu\nu}$ whose curvature scale ${\cal R}$ is much larger than the Schwarzschild radius.     Varying the In-In action in this background with respect to $x^\mu$ then yields the equation
\bea
\label{eq:dpds}
{D\over D s} p^\mu\equiv {dx^\rho\over ds}\nabla_\rho p^\mu =\langle Q^E_{\rho\sigma}\rangle \nabla^\mu E^{\rho\sigma} + \langle Q^B_{\rho\sigma}\rangle \nabla^\mu B^{\rho\sigma}, 
 \eea    
(this variation is done most conveniently in Gauss normal coordinates centered on the worldline, i.e. $\partial_\sigma g_{\mu\nu}(x(\lambda))  = 0$, and then covariantizing to obtain a result valid in any frame).    Using the general properties of the Schwinger-Keldysh generating function, the tidal moments (expectation values) $\langle Q^{E,B}_{\rho\sigma}(s)\rangle$ induced by the background curvature are given,  in the linearized approximation,  by
\beq
\label{eq:lresp}
\langle Q^{E}_{\mu\nu}(s)\rangle = \int ds^\prime G^{E;ret}_{\mu\nu;\rho\sigma}(s-s') E^{\rho\sigma}(x(s'))+{\cal O}(E^2),
\eeq
and similarly for $\langle Q^B_{\mu\nu}\rangle$, where the retarded Green's function of the operator $Q^{E/B}_{\mu\nu}$ is defined by
\beq
G^{E/B;ret}_{\mu\nu,\rho\sigma}(s-s') = - i \theta(s-s') \langle [Q^{E/B}_{\mu\nu}(s),Q^{E/B}_{\rho\sigma}(s')]\rangle,
\eeq
and the expectation value is  calculated at zero external field, $E_{\mu\nu}=B_{\mu\nu}=0$.    Thus Eq.~(\ref{eq:dpds}) describes the motion of a general compact object (not necessarily a black hole), in the limit of small radius, in terms of the response functions of moment operators.     These response functions in turn depend on the microphysics which describes the internal dynamics of the compact body.  At present we have no microscopic theory of these correlators, though we do know that they are constrained by sum rules~\cite{sumrules1,sumrules2}.

If the internal dynamics is fast compared to the time scale of the tidal perturbation, the  frequency space response functions  will be  a polynomial, with coefficients that depend on the internal structure of the compact object.   This is equivalent to the statement that the correlation functions die off exponentially fast at late times.  In~\cite{GnR2}, we showed that the first two terms in the low-frequency expansion of the retarded Green's function are related respectively to the object's tidal Love number and graviton absorption cross section.  The former/latter is conservative/dissipative and related  to the real/imaginary part of the frequency-space retarded Green's function.  For our present purposes we are only interested in the retarded propagator for a non-spinning (Schwarzschild) black hole,  for which the vanishing of the static response~\cite{Binnington:2009bb,Damour:2009va,Kol:2011vg} together with the results of~\cite{GnR2} fixes the response  to be of the form
\bea
\label{eq:VEVe}
\langle Q^{E}_{\mu\nu}(s)\rangle &=& {r^6_s\over 180 G_N}  \left(P_\mu{}^\rho P_\nu{}^\sigma-{1\over 3} P_{\mu\nu} P^{\rho\sigma}\right){ \dot E}_{\rho\sigma}(x(s))+\cdots , \\
\label{eq:VEVb}
 \langle Q^{B}_{\mu\nu}(s)\rangle &=& {r^6_s\over 180 G_N}  \left(P_\mu{}^\rho P_\nu{}^\sigma-{1\over 3} P_{\mu\nu} P^{\rho\sigma}\right){\dot B}_{\rho\sigma}(x(s))+\cdots ,
 \eea 
where $P^\mu{}_\nu = \delta^\mu{}_\nu - p^\mu p_\nu/p^2$, ${\dot E}_{\mu\nu} = {dx^\rho\over ds} \nabla_\rho E_{\mu\nu}$, ${\dot B}_{\mu\nu} = {dx^\rho\over ds} \nabla_\rho B_{\mu\nu}$, and terms with more time derivatives or more powers of curvature have been dropped.   The factors of the projection tensor $P_{\mu\nu}$ ensure that the expectation values are purely spatial and traceless in the rest frame of the black hole, so that that the response is purely quadrupolar.   Inserting these expectation values into Eq.~(\ref{eq:dpds}), the motion of a Schwarzschild black hole in a slowly varying background gravitational field becomes
\bea
\label{eq:dpdsbh}
{D\over D s} p^\mu \approx  {r^6_s\over 180 G_N}\left[{ \dot E}_{\rho\sigma} \nabla^\mu E^{\rho\sigma} +{\dot B}_{\rho\sigma} \nabla^\mu B^{\rho\sigma}\right].
 \eea  
 The right hand side is odd under time reversal $s\rightarrow -s$ as is characteristic of dissipative forces.   For example, using this result, we can calculate the rate of change of the black hole's mass as it moves through the tidal background, 
\beq
{\dot M} = {1\over  M} p\cdot {D\over D s} p \approx  {16 \over 45} (G_N^5 M^6) \left(\dot E_{\mu\nu} {\dot E}^{\mu\nu} + \dot B_{\mu\nu} {\dot B}^{\mu\nu}\right).
\eeq 
This is consistent with results obtained in~\cite{Poisson:2004cw} for the rate $\langle {\dot M}\rangle$ averaged over the motion of the black hole.

\section{Horizon dissipation in dynamical gravity}
\label{sec:2body}

The same worldline effective action formalism can also be applied to dissipation in dynamically generated spacetimes, i.e sourced by the compact objects, rather than the fixed background field case discussed above.   In order to do so, we have to include in the In-In functional an integral over the fluctuations\footnote{The role of the In-In formalism to describe radiation reaction forces was first discussed in~\cite{galley}.} of the gravitational field generated by the black hole sources.    In our applications below we will consider a binary system of black holes in two distinct kinematic regimes.   First, we will work out the effects of dissipation in the relativistic collision of two black holes at large impact parameter $b\gg r_s$, so that the system is amenable to a post-Minkowskian (PM) expansion.    As a second example, we will consider non-relativistic binaries and use our formalism to calculate the real-time (rather than time averaged) dissipative dynamics of a pair of comparable mass black holes in post-Newtonian (PN) orbits.

In both of these examples, we ignore the emission of on-shell gravitons.  Since the typical impact parameters and/or orbital distances are large compared to the internal scale $r_s$, the interactions between the black holes are mediated by purely ``potential'' graviton modes~\cite{GnR1} that do not go on-shell.    Thus, it is convenient to first integrate out these modes in order to obtain a (non-local in $x$) two-particle Lagrangian which can then be fed into the In-In generating function.      At leading order in a formal $G_N$ expansion, this yields a two-particle Lagrangian of the form
\beq
S[\chi_\alpha,e_\alpha,X_\alpha]=\sum_{\alpha=1}^2 S_{pp,\alpha} + S_{int},
\eeq
where $S_{pp,\alpha}$ is defined in (\ref{Spp}), and the second term $S_{int}$ is the interaction that is generated by integrating out potential graviton exchange.   As in the previous section,  $\chi_\alpha=(x_\alpha,p_\alpha)$.    

\begin{figure}[t]
    \centering
    \includegraphics[scale=0.22]{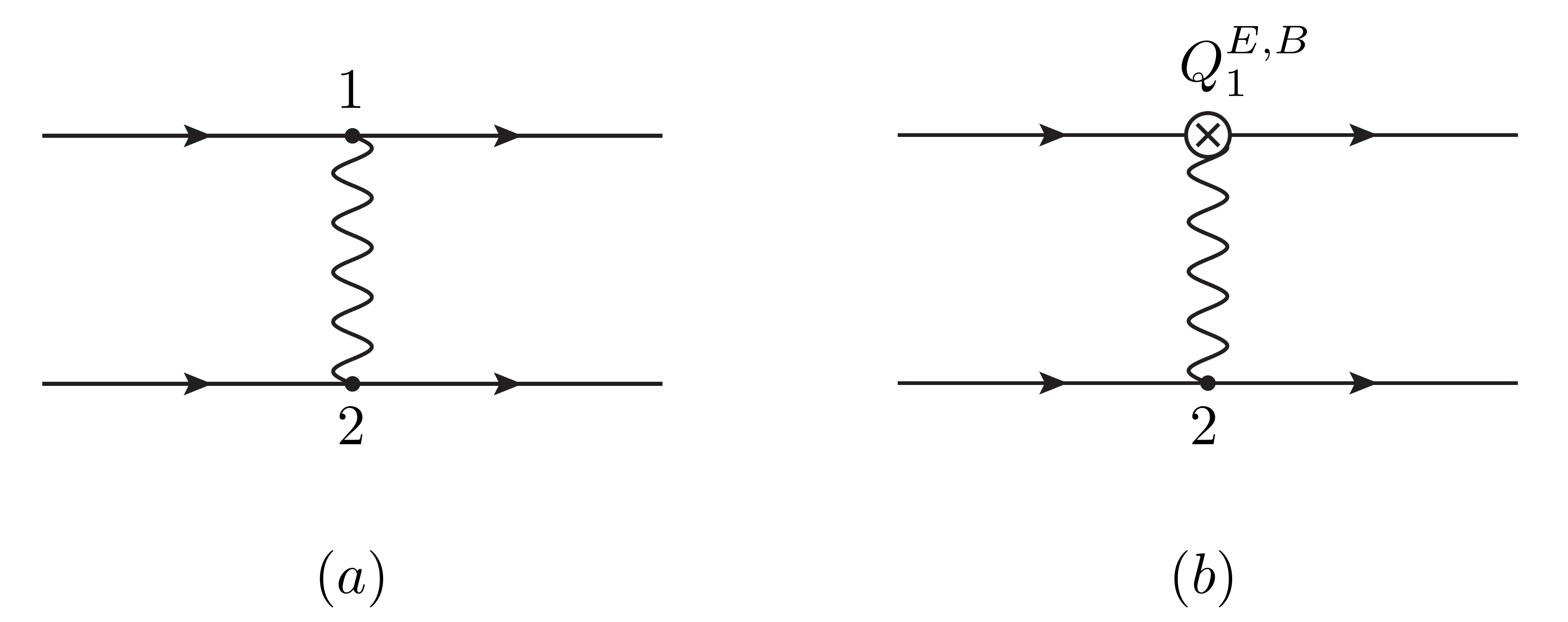}
\caption{Potential exchange diagrams that contribute to the two-particle action $S_{int}$.   In (a) the particles interact through the minimal gravitational interaction.    Figure (b) is the term in $S_{int}$ generated by the quadrupole couplings of particle 1 (a similar diagram with $1\leftrightarrow 2$ has been omitted).}
\label{fig}
\end{figure}

To integrate out the potential graviton modes, we fix deDonder gauge in which the graviton propagator has the form $D_{\mu\nu,\alpha\beta}(x) = P_{\mu\nu,\alpha\beta} D(x) $, where $P_{\mu\nu,\alpha\beta}={1\over 2}\left[\eta_{\mu\alpha}\eta_{\nu\beta}+\eta_{\mu\beta}\eta_{\nu\alpha}-\eta_{\mu\nu}\eta_{\alpha\beta}\right]$ and
\beq
D(x) = \int {d^4 k\over (2\pi)^4} {e^{i k\cdot x}}{i\over k^2}
\eeq
is the massless propagator.   The $i\epsilon$ prescription is not specified since we are in a kinematic regime in which the exchanged graviton is off-shell (the effects of radiation are higher order in the PM or PN power counting).    Then to leading order in powers of $G_N$ the gravitational interaction (neglecting internal structure) is given by Fig.~\ref{fig}(a),
\beq
\label{eq:losint}
S_{int} =   8\pi iG_N \int ds_1 ds_2 \left[(p_1\cdot p_2)^2 -{1\over 2} p_1^2 p_2^2\right] D(x_{12}) +\cdots.
\eeq
Here and in what follows we drop UV power-divergent self-interaction diagrams in which potential gravitons are emitted and re-absorbed by the same particle.    Such divergences may be  absorbed into the coefficients of local terms in the point particle action.

The internal dynamics mediated by the quadrupole operators generates an additional contribution to $S_{int}$, which to leading order in $G_N$ is given by the Feynman diagrams in Fig.~\ref{fig}(b) and takes the form  
\beq
\label{eq:dsint}
S_{int} =\cdots +   8\pi iG_N \sum_{\alpha=1,2}\int ds_1 ds_2 \left[Q^{E,\alpha}_{\mu\nu}(s_\alpha) T_{E,\alpha}{}^{\mu\nu,\rho\sigma} +Q^{B,\alpha}_{\mu\nu}(s_\alpha) T_{B,\alpha}{}^{\mu\nu,\rho\sigma} \right]\partial_\rho\partial_\sigma D(x_{12})+\cdots
%
%
%
\eeq
where we define
\bea
\nn
T_{E,1}{}^{\mu\nu,\rho\sigma}  &=& {1\over 2 p_1^2}\left[-\left(2 (p_1\cdot p_2)^2- p_1^2 p_2^2\right)\eta^{\mu(\rho} \eta^{\sigma)\nu} + p_2^2 \left(\eta^{\mu\nu} p_1^\rho p_1^\sigma- 2 p_1^{(\mu} \eta^{\nu)(\rho} p_1^{\sigma)}\right)
\right.\\
\label{eq:TE}
& & {}\left. \hspace{1cm}+ 4 (p_1\cdot p_2) p_1^{(\rho} \eta^{\sigma)(\mu} p_2^{\nu)} - 2 p_2^\mu p_2^\nu p_1^\rho p_1^\sigma\right],\\
\nn\\
\label{eq:TB}
T_{B,1}{}^{\mu\nu,\rho\sigma}  &=& {1\over 2 p_1^2}\left[(p_1\cdot p_2)\left( \epsilon^{\mu(\rho} \eta^{\sigma)\nu} + (\mu\rightarrow \nu)\right) - 2 p_2^{(\mu}\epsilon^{\nu)(\rho} p_1^{\sigma)}\right]
\eea
with $\epsilon^{\mu\nu} = \epsilon_{\mu\nu\rho\sigma} p_1^\rho p_2^\sigma$.  $T_{E,2}{}^{\mu\nu,\rho\sigma}$ and $T_{B,2}{}^{\mu\nu,\rho\sigma}$ are similarly defined by exchanging $p_1\leftrightarrow p_2$ in these equations.     Note that all variables appearing in Eqs.~(\ref{eq:dsint}),~(\ref{eq:TE}),~(\ref{eq:TB}) have arbitrary dependence on the worldline parameters $s_{1,2}$.    For example, $x_1^\mu=x^\mu_1(s_1),$ $p_1^\mu=p_1^\mu(s_1)$, etc..   Finally, we are ignoring spin, so there is no distinction between global (flat space) indices and local indices in these expressions.

\subsection{Inelastic BH-BH scattering in the PM approximation}
\label{sec:PM}

We consider first the inelastic collision $1+2\rightarrow 1^*+2^*$ of two black holes with no gravitational radiation produced in the final state.  For the initial configuration, we take a pair of black holes with asymptotic worldlines $x^\mu_{\alpha=1,2}(s\rightarrow -\infty) = p_\alpha^\mu s + b_\alpha^\mu$.  Without loss of generality, the impact parameter $b^\mu=b_1^\mu-b_2^\mu$  can be taken to be orthogonal to the initial momenta, $p_{1,2}\cdot b =0$, and spacelike, with magnitude squared $b_\mu b^\mu = -b^2 <0$.   We work in the Post-Minkowskian limit $G_N E/b\ll 1$, where $E$ is the typical energy in the scattering process.    

In the inelastic scattering process, the final state black holes $1^*$ and $2^*$ have invariant masses $m_{1^*,2^*}$ which differ from those of the initial states, with mass $m_{1,2}$.    We therefore split up the momentum transfer to a given black hole into elastic and inelastic parts,
\beq
\Delta p^\mu= \Delta p^\mu_{el} + \Delta p^\mu_{in}.
\eeq
The term $\Delta p^\mu_{el}$ corresponds to the momentum transfer neglecting dissipative finite size effects, but including in principle all-orders PM corrections due to potential graviton exchange and radiation reaction.   Such effects do not change the invariant mass of the asymptotic black hole states and therefore $\Delta p^\mu_{el}$ obeys a constraint
$(p+ \Delta p_{el})^2=p^2$.   In the PM approximation, we expect that the inelastic part  $\Delta p^\mu_{in}\ll \Delta p^\mu_{el}$, so that the change in the black hole masses is $\Delta m^2\approx 2 p\cdot \Delta p_{in}$.   The total deflection in the four-momentum of each black hole caused by the collision is related to the In-In action by\footnote{We denote $\int ds = \int_{-\infty}^{\infty} ds$.}
\beq
\Delta p_{\mu,\alpha} = \int ds {d\over ds} p_{\mu,\alpha}(s)= \int  ds {\delta \over \delta x^\mu_\alpha(s)} S_{0,\alpha} = -\int ds\left\langle  {\delta \over \delta x^\mu_\alpha(s)} S_{int}[\chi,X] \right\rangle.
\eeq

As a warm up exercise, consider  the case of elastic scattering at leading order in $G_N$. From Eq.~(\ref{eq:losint}), we have
\beq
\int ds\left\langle  {\delta \over \delta x^\mu{}_1(s)} S_{int}\right\rangle = 8\pi i G_N \int ds_1 ds_2 \left[(p_1\cdot p_2)^2 -{1\over 2} p_1^2 p_2^2\right] {\partial\over\partial x^\mu_1}D(x_{12}).
 \eeq
To leading order in perturbation theory, we may take the momenta on the LHS to be constants, and the trajectories $x_{1,2}$ to be undeflected free particle paths.  Thus,
\beq
\int  ds_1 ds_2 {\partial\over\partial x^\mu{}_1} D(x_{12}) = {\partial\over\partial b^\mu} \int  ds_1 ds_2 D(x_{12}) 
\eeq
and using the integral
\beq
 \int  ds_1 ds_2 D(x_{12}) = \int {d^4 k \over (2\pi)^4}  (2\pi) \delta(k\cdot p_1) (2\pi)\delta(k\cdot p_2) {e^{-i k\cdot b}\over k^2}= {i\over 4\pi} {1\over\sqrt{(p_1\cdot p_2)^2 - p_1^2 p_2^2}} \ln\left(-\mu_{IR}^2 b_\mu b^\mu\right),
 \eeq
 (we have introduced an arbitrary infrared cutoff $\mu_{IR}$ to make sense of the divergent momentum integral) we recover the standard result~\cite{Westpfahl:1979gu,Portilla:1980uz} 
 \beq
 \label{eq:1PM}
\Delta p_1^\mu = -\Delta p_2^\mu =  -{4 G_N m_1 m_2\over b^2}  {(v_1\cdot v_2)^2 -{1\over 2}\over \sqrt{(v_1\cdot v_2)^2 - 1}} {b^\mu},
\eeq 
written in terms of the initial four-velocities, $v^\mu=p^\mu/\sqrt{p^2}$.

The inelastic part of the momentum deflection involves the expectation value of the multipole operators,
\bea
\nn
\Delta p_{1,in}^\mu &=& - 8\pi iG_N \sum_{\alpha=1,2}\int ds_1 ds_2 \left[\langle Q^{E,\alpha}_{\lambda\nu}(s_\alpha)\rangle T_{E,\alpha}{}^{\lambda\nu,\rho\sigma} +\langle Q^{B,\alpha}_{\lambda\nu}(s_\alpha)\rangle T_{B,\alpha}{}^{\lambda\nu,\rho\sigma} \right]{\partial\over\partial x_1^\mu} \partial_\rho\partial_\sigma D(x_{12}).
\eea
Here, the expectation values are the moments induced by the tidal interactions between the sources given in Eq.~(\ref{eq:dsint}).   In the linear response approximation, these can be read off Eq.~(\ref{eq:lresp}).  For instance,
\beq
\label{eq:qevev}
 \left\langle {Q_{E1}^{\mu\nu}(s)}\right\rangle =  \int ds_{1} G_{E1,ret}{}^{\mu\nu,\rho\sigma}(s-s_{1})\left[-8\pi i G_N T_{E,1}^{\mu\nu,\rho\sigma} \partial_\rho\partial_\sigma D(x_{12})\right],
 \eeq
 where the part inside the square brackets represents the linearized tidal electric field produced by black hole 2.     Using the retarded Green's function from Eq.~(\ref{eq:VEVe}), we have 
  \beq
    \left\langle {Q_{E1}^{\mu\nu}(s)}\right\rangle \approx -{32 G_N^6 p_1^4\over 45} \left[T_{E,1}^{\mu\nu,\rho\sigma} {d\over ds} I_{\rho\sigma}(p_1 s + b,p_2)\right]_{TT,1},
 \eeq
 and 
 \beq
    \left\langle {Q_{E2}^{\mu\nu}(s)}\right\rangle \approx -{32 G_N^6 p_2^4\over 45} \left[T_{E,2}^{\mu\nu,\rho\sigma} {d\over ds} I_{\rho\sigma}(p_2 s - b,p_1)\right]_{TT,2},
 \eeq
 and similarly, using Eq.~(\ref{eq:VEVb}),  the magnetic expectation values take the same form.   In these expressions $TT,\alpha=1,2$ denotes the transverse to $p_\alpha$ traceless projection of the tensor defined in Eqs.~(\ref{eq:VEVe})~(\ref{eq:VEVb}).    We have also used the result
 \beq
\int ds D(x-p s) = -{i\over 4\pi} {1\over\sqrt{(p\cdot x)^2 - x^2 p^2}}
\eeq
and defined the tensor 
\beq
I_{\mu_1\cdots \mu_n}(x,p) = {\partial\over\partial x^{\mu_1}} \cdots  {\partial\over\partial x^{\mu_n}}  {1\over \sqrt{(p\cdot x)^2 - x^2 p^2}}.
\eeq

Using these intermediate results, the momentum transfer can be written as an integral over a single worldline parameter, 
\bea
\nn
\Delta p_{1,in}^\mu =& -2 G_N\int ds I^{\mu\lambda\nu}(p_1 s + b,p2) \left[T_{E,1}^{\rho\sigma,\lambda\nu}     \left\langle {Q_{E1,\rho\sigma}(s)}\right\rangle +T_{B,1}^{\rho\sigma,\lambda\nu}     \left\langle {Q_{B1,\rho\sigma}(s)}\right\rangle \right.\\
\nn\\
&\left. {} \hspace{3.5cm}- \left(p_1\leftrightarrow p_2, b^\mu\leftrightarrow -b^\mu\right)\right]
\eea
The first two terms in this equation correspond to the self-force induced by the dynamical moments of particle 1 on its own motion, while the second line gives the transfer of particle 1 energy-momentum into the internal degrees of freedom of the second black hole.    We have performed the tensor contractions using FeynCalc~\cite{fc}, while the parameter integration reduces to elementary integrals of the form
\beq
\int ds {s^n\over [s^2+1]^m} = {1+(-1)^n\over 2}\cdot {\Gamma\left({n+1\over 2}\right) \Gamma\left(m-{n+1\over 2}\right)\over \Gamma(m)}.
\eeq
The result is then  $\Delta p_{2,in}^\mu =- \Delta p^\mu_{1,in}$, where
\bea
\nn
\Delta p^\mu_{1,in} &=& -{5\pi\over 32} {G_N^7 m^4_1 m^4_2\over b^7} \left[{21 (v_1\cdot v_2)^4 - 14 (v_1\cdot v_2)^2 +9\over \sqrt{(v_1\cdot v_2)^2 -1}}\right.\\
\nn
& &\left. {} +{7} (1+ 3(v_1\cdot v_2)^2)  \sqrt{(v_1\cdot v_2)^2 -1}\right]\left[{m_1^2\over m_2^2} (v_1-(v_1\cdot v_2) v_2)^\mu-(1\leftrightarrow 2)\right]\\
\nn
\\
\nn\\
\label{eq:result}
&=&  -{5\pi\over 16} {G_N^7 m^4_1 m^4_2\over b^7} {P(v_1\cdot v_2)\over \sqrt{(v_1\cdot v_2)^2 -1}} \left[{m_1^2\over m_2^2} (v_1-(v_1\cdot v_2) v_2)^\mu-(1\leftrightarrow 2)\right],
\eea
with $P(\gamma) = 21 \gamma^4 -14 \gamma^2 + 1$.  In the first equality we have split up the result into an electric contribution on the first line and the magnetic part on the second, while in the second equality we present the combined result.   As a simple non-trivial check of this formula, we find that the change in the invariant mass of the black hole as a result of the collision
\beq
\label{eq:dm2}
\Delta p_1^2 \approx 2 p_1\cdot \Delta p_{1,in} = {5\pi\over 16} {G_N^7 m^7_1 m^2_2\over (b^2)^{7/2}} P(\gamma) \sqrt{\gamma^2-1} >0
\eeq
is positive over the entire kinematic range $\gamma=v_1\cdot v_2 \geq 1$, and likewise $\Delta m_2^2>0$.   Given that the change in entropy (horizon area) of a pair of black holes that scatter in from and out to infinity is simply $\Delta S = 4\pi G_N \left(\Delta m_1^2+\Delta m_2^2\right)>0$, our result is found to be consistent with Hawking's area theorem~\cite{Hawking:1971vc}.

Similarly, we can compute the differential distribution of final state black hole masses in the scattering process $1+2\rightarrow 1^*+2^*$.   In the point-particle approximation (i.e. neglecting tidal dissipation at the horizon), the masses of the colliding black holes are unchanged to any order in PM perturbation theory, $m_{1,2}^*=m_{1,2}$, even including the effects of radiation emission or backreaction.   The leading order cross section $d\sigma/d\Delta m_1^2,$ with $\Delta m_1^2=m_{1^*}^2-m_1^2\ll m_{1,2}^2$ instead follows from Eq.~(\ref{eq:dm2}).    For instance, in the high energy limit $\gamma\gg 1$, this observable takes the form\footnote{Equivalently, in terms of the horizon area (in the asymptotic rest frame of each black hole)  $A=16\pi (G_N m)^2$, the distribution reads ${1\over A}_1 \cdot {d\sigma\over d \log \Delta A_1} \approx  {1\over 7} \left[{(105\pi)^2\over 2^{29}}\right]^{1/7}\gamma^{10/7} \left[{A_2\over\Delta A_1}\right]^{2/7}$, up to corrections whose relative size is of order $(\Delta A/A)^{1/7}$.    The differential distribution in terms of geometric quantites is independent of $G_N$, as expected given that we are dealing with solutions of the vacuum Einstein equations $R_{\mu\nu}=0$.}
\beq
{d\sigma\over d \Delta m_1^2} = 2 \pi\left|b {\partial b\over\partial\Delta m_1^2}\right| \approx \left[{225\pi^9\over 33614}\right]^{1/7} (G_N m_1)^2 \gamma^{10/7} m_2^{4/7} (\Delta m_1^2)^{-9/7},
\eeq
up to fractional errors of order $(\Delta m_1^2/m_{1,2}^2)^{1/7}\ll 1$.   The overall numerical factor and the non-analytic dependence on the kinematic quantities $\gamma,\Delta m^2$ are non-trivial predictions of classical GR which could in principle be tested against numerical simulations of black hole collisions at finite impact parameter.

Our result in Eq.~(\ref{eq:result}) can also be used to calculate dissipative effects in PM scattering.   A quantity that has received attention in the recent literature~\cite{Damour:2016gwp,Damour:2017zjx,Bini:2018ywr,Antonelli:2019ytb,Damour:2019lcq,Bini:2020flp}, due to its connection with the EOB framework~\cite{EOB} for parameterizing gravitational wave templates, is the scattering angle $\chi(s,b)$ defined in the center-of-mass frame ${\vec p}\equiv {\vec p}_1=-{\vec p}_2$.     Our result in Eq.~(\ref{eq:result}) then contributes an amount\footnote{Note that in terms of the  CM energy $E_{CM}=\sqrt{\hat s}$, we have $\gamma(\hat s)=\left. (\hat s^2 -m_1^2 - m_2^2)\right/ m_1 m_2$, and CM frame momentum $|{\vec p}|=m_1 m_2 \sqrt{\gamma^2-1}/\sqrt{\hat s}$.  The and orbital angular momentum in the CM frame is $J=|{\vec p}| b$.}   
\beq
\label{eq:8PM}
{\Delta \chi_{in}\over\chi_{1PM}}= {5\pi\over 16} \left({G_N m_1 m_2\over J}\right)^7 P(\gamma(\hat s)) \left[{1\over m_1^3}+{1\over m^3_2} + \gamma(\hat s) \left({m_1^5+m_2^5\over m^4_1 m_2^4}\right)\right] {|{\vec p}|^6\over\sqrt{\hat s}} 
\eeq
relative to the leading order result $\chi_{1PM}(\hat s,b)$ resulting from Eq.~(\ref{eq:1PM}).

 \subsection{Post-Newtonian equations of motion for binary dynamics}
 \label{sec:PN}

 As a second application, we compute the damping forces due to the presence of the horizon in the case of a binary black hole in a non-relativistic orbit.   For non-relativistic particles, we have $D(x) \approx -{i\over 4\pi |{\vec x}|} \delta(x_0)$, and the two-particle interaction term in Eq.~(\ref{eq:dsint}) is dominated by the electric tidal interaction, which reduces to  
 \beq
 S_{int} \approx -G_N m_1 m_2 \int dt  \left({Q^{ij}_{E,1}(t)\over m_1^2} + {Q^{ij}_{E,2}(t)\over m_2^2}\right) \partial_i \partial_j {1\over |{\vec x}(t)|},
 \eeq
 with ${\vec x}={\vec x}_1-{\vec x}_2$, up to terms suppressed by more power of the velocities.   Varying the In-In action, we obtain in the linear response limit, an instantaneous non-conservative force on the black holes that is given by,
\bea
\nn
{\vec F}_1(t) &=& {\delta\over \delta {\vec x}_1(t)}\left. \Gamma[{\vec x},\tilde{\vec x}]\right|_{\vec x=\tilde{\vec x}} = -G_N m_1 m_2  \left\langle {Q^{jk}_{E,1}(t)\over m_1^2} + {Q^{jk}_{E,2}(t)\over m_2^2}\right\rangle\nabla \partial_j \partial_k {1\over |{\vec x}(t)|} = -{\vec F}_2(t),\\
\eea
with ${\vec x}={\vec x}_1-{\vec x}_2$. The In-In expectation values in the relativistic limit can be obtained from Eq.~(\ref{eq:qevev}),
\bea
\langle Q^{ij}_{E,1}\rangle(t) &=& {16\over 45} G_N^6 m_1^7 m_2 {d\over dt} \partial^i \partial^j |{\vec x}(t)|^{-1},\\
\eea
and similarly for $\langle Q^{ij}_{E,2}\rangle(t)$.   The force can then be expressed as
\bea
{\vec F}_1(t)= -{\vec F}_2(t) = -{32\over 5} {G_N^7 (m_1 m_2)^2 (m_1^4 + m_2^4)\over |{\vec x}|^8} \left[{\vec v} +  {2{\vec v} \cdot {\vec x}_{12}\over |{\vec x}|^2} {\vec x}\right],
\eea
where and ${\vec v}= {d\over dt} {\vec x}$ is the relative velocity.   This friction force is a 6.5PN effect, analogous to the Burke-Thorne~\cite{BT} potential which encapsulates leading order radiation reaction effects on the orbits, arising at 2.5PN order. 

As a simple example of this result, we compute the instantaneous (as opposed to time averaged) mechanical power in a binary that is converted into mass by the horizons of the two black holes ($M=m_1+m_2$, $\mu = m_1 m_2/M$)
\beq
\label{eq:nw}
{d\over dt } E_{h} = \sum^2_{\alpha=1} {\vec v}_\alpha \cdot {\vec F}_\alpha = -{32\over 5} G_N^{-1} \left({G_N M\over |{\vec x}|}\right)^8 \left({\mu\over M}\right)^2 \left(1 - {4\mu\over M} + {2 \mu^2\over M^2}\right) \left[{\vec v}^2 + {2 ({\vec v}\cdot {\vec x})^2\over |{\vec x}|^2}\right],
\eeq
which agrees with and generalizes the result of~\cite{Poisson:1994yf,Tagoshi:1997jy} valid in the limit $\mu\ll M$ where black hole perturbation theory holds.   Similarly, the decay of the CM frame orbital angular momentum is given by
\beq
{d\over dt} {\vec L}_{CM} = \sum^2_{\alpha=1} {\vec v}_\alpha \times {\vec F}_\alpha = {64\over 5} G_N^7 \mu M^6 \left(1 - {4\mu\over M} + {2 \mu^2\over M^2}\right) {{\vec v}\cdot {\vec x}\over |{\vec x}|^{10}} {\vec L}_{CM}.
\eeq

\section{Conclusions}
\label{sec:conc}

The inelastic dynamics of compact astrophysical objects are of particular interest  due to their dependence on the internal structure of
the body. In general this internal structure depends on non-gravitational physics (e.g. equation of state), but in the case of a black
hole it is possible to calculate in a model independent way, as the vacuum Einstein equations are sufficient to determine all of the dynamics. 
 In this paper we have calculated the effects of dissipation on the instantaneous equations of motion for black hole binary systems in the PN and PM limits.   In particular, we have extended our results in~\cite{GnR2}, where only the time averaged dissipation was discussed\footnote{We have also corrected the PN results in~\cite{GnR2} for the dissipated power.   Eq.~(\ref{eq:nw}) above replaces the incorrect Eq.~(36) of~\cite{GnR2}.}.   These results complement the calculations in~\cite{Bini:2020flp,Cheung:2020sdj} for the PM scattering angle resulting from conservative finite size effects.

In order to treat dissipation in the case of more general compact objects, the only modifications needed to the calculations presented in this paper are the retarded correlators which we define in Eqs.~(\ref{eq:VEVe}),~(\ref{eq:VEVb}).    On general grounds, in the frequency domain, the imaginary part (i.e. the part responsible for absorption) of these response functions would still be linear at small frequency, but with a numerical coefficient that differs from the one for black holes.   If this coefficient scales with radius in the same way as for black holes, as $R^6/G_N$, this could yield an enhancement of dissipative effects relative to naive PN power counting, similar to the enhancement of conservative tidal effects discussed in~\cite{Flanagan:2007ix,Hinderer:2007mb}.   One would also expect enhanced dissipation for maximally rotating black holes, analogous to the enhancement from 4PN to 2.5PN for the energy loss in extreme mass ratio black hole inspirals found in~\cite{Poisson:1994yf,Tagoshi:1997jy} (which is, ultimately, a consequence of superradiance~\cite{staro}, as discussed in~\cite{Porto:2007qi,Endlich:2016jgc}).   We hope to explore some of these directions in future work.

\section{Acknowledgments}

This work was partially supported by the US Department of Energy under grants DE-SC00-17660 (WG) and DE- FG02-04ER41338 and FG02- 06ER41449 (IZR).

\end{document}